\documentclass[preprint, aps, amsmath,amssymb]{revtex4-1}
\usepackage{graphicx}

\pdfoutput=1
\usepackage{amsmath}
\usepackage{color}
\usepackage{float}
\usepackage{lipsum}

\begin{document}
\title{Unconventional Superconductivity at Mesoscopic Point-contacts on the 3-Dimensional Dirac Semi-metal Cd$_3$As$_2$}

\author{Leena Aggarwal$^1$}

\thanks{These authors contributed equally to the work}
\author{Abhishek Gaurav$^1$}
\thanks{These authors contributed equally to the work}

\author{Gohil S. Thakur$^2$}

\author{Zeba Haque$^2$}


\author{Ashok K. Ganguli$^{2,3}$}
\email{ashok@chemistry.iitd.ac.in}

\author{Goutam Sheet$^1$}
\email{goutam@iisermohali.ac.in}
\affiliation{$^1$Department of Physical Sciences,  
Indian Institute of Science Education and Research Mohali,
Sector 81, S. A. S. Nagar, Manauli, PO: 140306, India}

\affiliation{$^2$Department of Chemistry, Indian Institute of Technology, New Delhi
110016, India}

\affiliation{$^3$Institute of Nano Science \& Technology, Mohali 160064, India}

\begin{abstract}

Since the three dimensional (3D) Dirac semi-metal Cd$_3$As$_2$ exists close to topological phase boundaries, in principle it should be possible to drive it into exotic new phases, like topological superconductors, by breaking certain symmetries. Here we show that the mesoscopic point-contacts between silver (Ag) and Cd$_3$As$_2$ exhibit superconductivity up to a critical temperature (onset) of 6 K while neither Cd$_3$As$_2$ nor Ag are superconductors. A gap amplitude of 6.5 meV is measured spectroscopically in this phase that varies weakly with temperature and survives up to a remarkably high temperature of 13 K indicating the presence of a robust normal-state pseudogap. The observations indicate the emergence of a new unconventional superconducting phase that exists only in a quantum mechanically confined region under a point-contact between a Dirac semi-metal and a normal metal.
\end{abstract}

\maketitle

Although the 3D topological Dirac semi-metals\cite{LiuNa3Bi, LiuCd3As2, YoungPRL} can be stable where the Dirac points are protected by symmetry, they exist close to topological phase boundaries, and therefore, they are expected to be driven into topologically distinct phases by explicit symmetry breaking\cite{WangPRB}. The proposed new phases of matter that could be derived from a 3D topological Dirac semi-metal through explicit symmetry breaking include new topological insulators\cite{TI}, Weyl semi-metals\cite{BurkovPRL, Waniridate}, and topological superconductors\cite{TSReview, TSPRL, WangPRB2, QiPRB}. Therefore, the 3D Dirac semi-metals provide a novel platform where topologically diverse quantum phase transitions can be realized. Recently the II-V semiconductor Cd$_3$As$_2$ has been shown to be a stable 3D Dirac semi-metal by angle resolved photoemission spectroscopy (ARPES)\cite{LiuCd3As2} and scanning tunneling spectroscopy (STS)\cite{YazdaniSTM} experiments. The ARPES experiments concluded that Dirac fermions exist in the bulk of Cd$_3$As$_2$ with linear energy dispersion along all three directions in the momentum space. The STS experiments revealed that the impurities in Cd$_3$As$_2$ influence only the valence bands which explains remarkably high mobility in Cd$_3$As$_2$.

All the theoretical and experimental works mentioned above have emphasized on one key point--that it should be possible to drive Cd$_3$As$_2$ into even more exotic states of matter and devices with novel functionalities can be obtained on Cd$_3$As$_2$. However, until now a practical realization of these ideas have been lacking. In this report, we show that nanometer scale junctions between pure elemental normal metals like silver (Ag), platinum (Pt) and gold (Au) and the 3D Dirac semi-metal Cd$_3$As$_2$ exhibit unconventional superconductivity up to a critical temperature $T_c$ of 6 K. A prominent, nearly temperature-independent pseudogap is observed up to a remarkably high temperature of 13 K.

\begin{figure*}[!h]
\includegraphics[width=\textwidth]{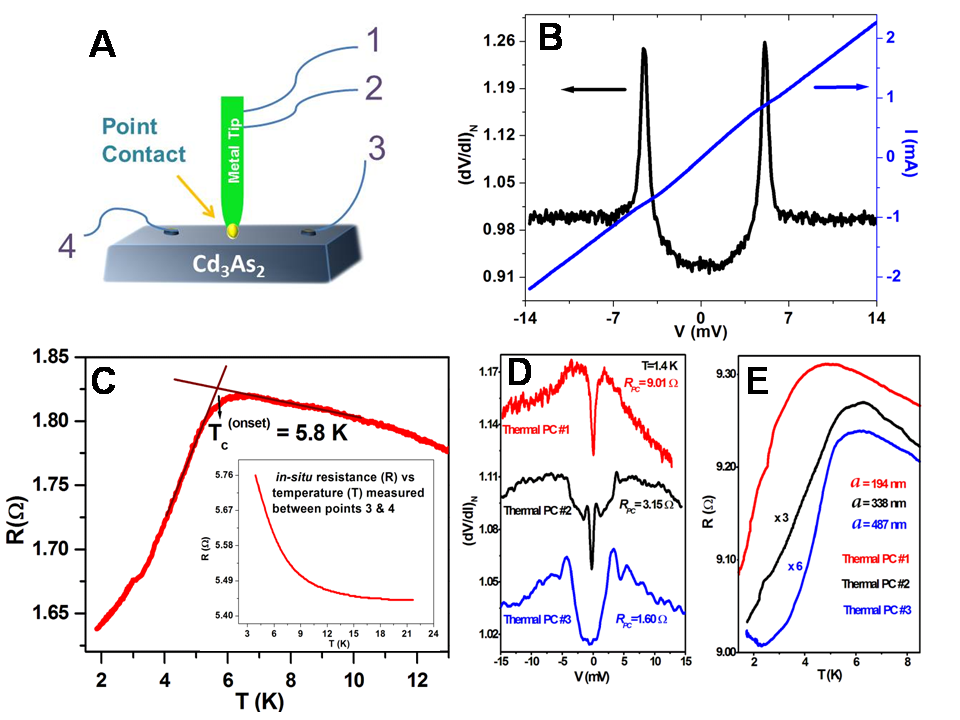}
\caption{(A) The schematic diagram representing the point-contact and the measurement electrodes. (B) A normalized differential resistance ($(dV/dI)_N$) spectrum in the thermal regime: The peaks in $(dV/dI)_N$ originate due to the critical current of the superconducting junction. (C) Resistance ($R$) of the point-contact as in (B) vs. temperature ($T$) showing the superconducting transition. The inset depicts the two probe resistance of the sample measured between the electrodes 3 and 4 {\textit{in-situ}}. The sample does not show superconducting transition confirming that the superconductivity occurs only under the point-contact. (D) Additional spectra in the thermal regime showing the superconductivity-related features such as drop in differential resistance at lower voltage and the critical current driven peaks in $(dV/dI)_N$ and (E) R-T curves for corresponding point-contacts in (D) showing superconducting transition. 
 }
\end{figure*}

In Figure 1 (A) we demonstrate a schematic diagram describing how the mesoscopic point-contacts between the metallic tips and Cd$_3$As$_2$ were formed {\textit{in-situ}} at low temperatures and how the transport and spectroscopic properties were measured that lead to the discovery of the superconducting phase\cite{Supplementary}. We first brought a sharp silver tip in contact with the sample (Cd$_3$As$_2$) to form a point-contact and measured the differential resistance ($dV/dI$) of the point-contact in a pseudo-four-probe geometry by a lock-in modulation technique \cite{Supplementary}. 

The resistance $R_{PC}$ of such a point-contact is generally given by Wexler's formula\cite{Wexler}: $R_{PC} = \frac{2h/e^2}{(ak_F)^2} + \Gamma (l/a)\frac{\rho (T)}{2a}$, where $h$ is Planck's constant, $e$ is the charge of a single electron, $a$ is the contact diameter, $\Gamma(l/a)$ is a slowly varying function of the order of unity, $\rho$ is the bulk resistivity of the material and $T$ is the effective temperature at the point-contact. The first term is known as Sharvin's resistance ($R_S$)\cite{SupriyoDatta} that depends on the fundamental constants $h$, $e$ and the number of conducting channels. Therefore $R_S$ is temperature independent. The second term is called the Maxwell's resistance ($R_M$) which depends directly on the resistivity of the materials forming the point-contact. The above equation also suggests that when the contact diameter is small (i.e., the contact is in the ballistic regime where the contact diameter is smaller than the inelastic and elastic mean free paths), $R_S$ decides the contact resistance that remains temperature independent. However, when the contact diameter is large (thermal regime), $R_M$ contributes most to the total contact resistance. Therefore resistive transitions lead to non-linearities in the $I-V$ characteristics of point-contacts in the thermal regime\cite{GoutamPRB}. No spectroscopy can be performed in the thermal regime due to the permitted inelastic processes at the interface. However, the point-contacts in the thermal regime can be used to detect the existence of superconductivity in complex systems\cite{GoutamPRB}. Two sharp peaks symmetric about $V$ = 0 in $dV/dI$ vs. $V$ plots obtained from thermal limit point-contacts is a signature of superconductivity in the point-contacts\cite{GoutamPRB, NaidyukBook}.

In this report we present the data obtained from Cd$_3$As$_2$ point-contacts with Ag tips in detail. However, we have observed similar results with Pt and Au tips. Please see supplementary material for the data obtained with other metallic tips\cite{Supplementary}.

\begin{figure*}[!h]
\includegraphics[width=\textwidth]{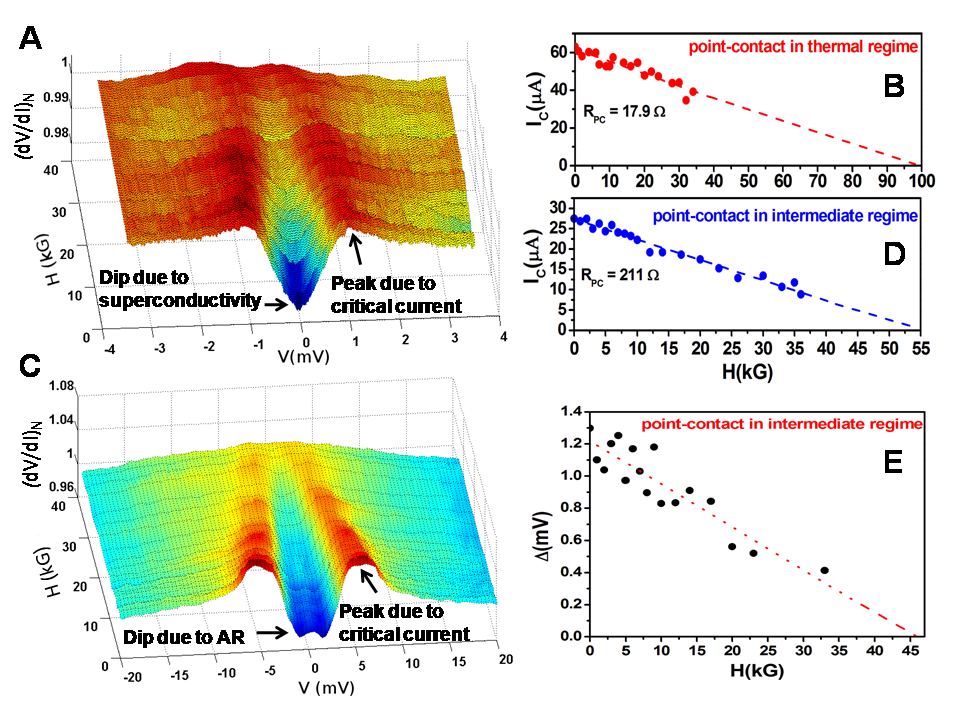}
\caption{(A) The magnetic field dependence of a spectrum in the thermal regime. The peaks in normalized differential resistance ($(dV/dI)_N$) due to critical current of the superconducting point-contact are indicated by arrows. (B) Magnetic field dependence of critical current ($I_C$) for the point-contact in the thermal regime. (C) The magnetic field dependence of a point-contact spectrum in the intermediate regime. The peaks in $(dV/dI)_N$ originating from critical current of the superconducting point-contact and the dips in $(dV/dI)_N$ originating from the Andreev reflection (AR) at the interface between Cd$_3$As$_2$ and Ag are indicated by arrows. (D) The magnetic field dependence of the critical current for the spectrum in the intermediate regime.  (E) Magnetic field dependence of the position of the AR dip (this is approximately equal to the gap $\Delta$. However, it is known that in the non-ballistic limit the gap is always underestimated\cite{GoutamPRB}). The dotted lines are linear fits.}
\end{figure*}

The measured normalized differential resistance ($(dV/dI)_N$) as a function of a dc-bias ($V$) developed across a point-contact between Ag and Cd$_3$As$_2$ is shown in Figure 1 (B). Two sharp peaks in $(dV/dI)_N$ are observed at $\pm 5 mV$ respectively. This spectrum is strikingly similar to the typical $(dV/dI)_N$ spectra obtained from a superconducting point-contact when the contact is close to the thermal regime of transport as discussed above. The peaks appear when the dc-current flowing through the point-contact reaches the critical current for the given point-contact. Below that voltage $(dV/dI)_N$ shows a dip due to the transition to the superconducting state\cite{GoutamPRB}. We have also measured the resistance of the point-contact (R) as a function of temperature (T) by sending an ac-current through electrodes 1 and 4 and by measuring the ac-voltage drop between the electrodes 2 and 3 with applied $V$ = 0. As shown in Figure 1(C), we clearly see the signature of a superconducting transition in the R-T plot. The onset of this transition is at 6K and the transition completes at 1.4 K with a broad transition width of $\sim$ 5 K \cite{contact}. Since neither Cd$_3$As$_2$ nor Ag are superconductors\cite{Supplementary} it is rational to conclude that only a confined region at the interface between Ag and Cd$_3$As$_2$ becomes superconducting giving rise to a novel state of matter that exists only in a confined region involving a 3-D Dirac semi-metal.

The critical current of a point-contact in the thermal limit should be dependent on the contact geometry and therefore the over-all shape of the $(dV/dI)_N$ spectrum and the position of the critical current driven peaks in $(dV/dI)_N$ for superconducting point-contacts must vary from contact to contact. In order to confirm this fact with the point-contacts formed on Cd$_3$As$_2$, we have repeated the measurements on more than 50 point-contacts on two samples grown in different batches. Some of the representative $(dV/dI)_N$ spectra are shown in Figure 1(D) and the corresponding R-T data are also shown in Figure 1(E). For all the point-contacts in the thermal limit we observe the critical current driven peaks in $(dV/dI)_N$ and the position of the peaks are different for different contacts confirming the geometry dependence of critical current. The size of the point-contacts that we have investigated varied between 20 nm and 450 nm (estimated using Wexler's formula)\cite{Supplementary}. No systematic dependence of the critical current on the contact size was found.

\begin{figure*}[!h]
\includegraphics[width=\textwidth]{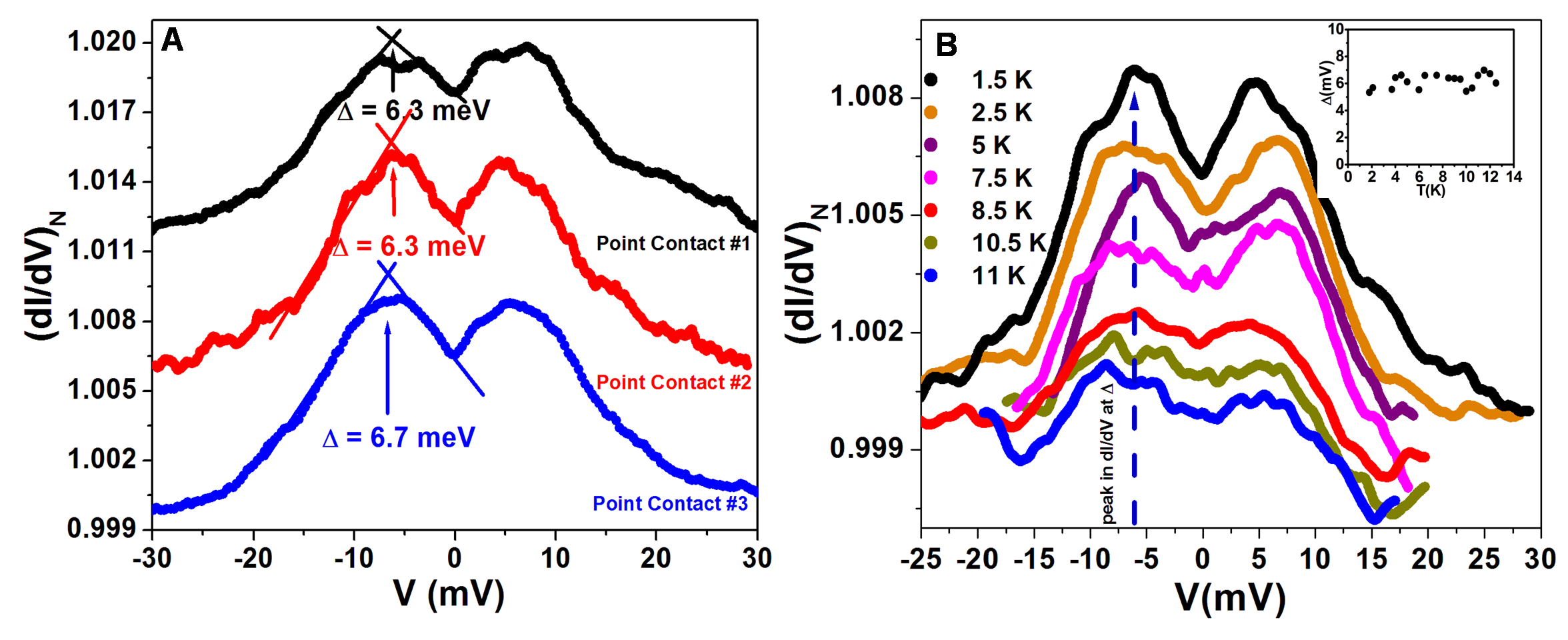}
\caption{(A) Spectra captured from three point-contacts in the ballistic regime showing the gap ($\Delta$) features. The spectra have been given a vertical shift for visual clarity. The scheme of computing $\Delta$ is also shown. (B) Temperature evolution of a spectrum showing the gap feature ($\Delta$). {\textit{inset:}} The temperature dependence of $\Delta$. In this plot we show more data points than the number of spectra shown. $\Delta$ survives above 13 K which is more than 2 times $T_c$ and does not show considerable temperature dependence indicating the presence of a pseudogap.
 }
\end{figure*}
In Figure 2(A) we show how the critical current dominated spectra obtained from the point-contacts in the thermal limit evolve with magnetic field. The central dip (due to superconducting transition below the critical current) in the spectra evolves smoothly with magnetic field and disappear at $H_c$ = 4.5 T. The critical-current driven side-peaks also show systematic evolution with magnetic field. Since the position of the peaks directly give the value of the critical current for a given point-contact, it is possible to plot the critical current as a function of field directly from this data. This plot has been shown in Figure 2(C). As expected for superconducting junctions, the critical current decreases with increasing magnetic field\cite{critical field}.   

When the point-contacts are in the ballistic regime of transport, where the quasiparticles do not undergo scattering within the contact region, it is possible to perform Andreev reflection spectroscopy and energy resolved information about the superconducting energy gap can be extracted\cite{BTK}. In order to obtain point-contacts in the ballistic limit we have followed the recipe described by Sheet\textit{et al.}\cite{GoutamPRB}. Following this recipe, we have withdrawn the Ag tip slowly until the critical current driven peaks disappeared and the features associated with Andreev reflection appeared in the spectra. As the tip is withdrawn slowly, the thermal regime point-contacts transition to a ballistic regime through an intermediate regime where both $R_S$ and $R_M$ contribute\cite{GoutamPRB}. One representative spectrum in the intermediate regime is shown in Figure 2 (C). As expected for superconducting point-contacts in the intermediate regime of transport, we have found signature of critical current driven features (peaks in $(dV/dI)_N$ symmetric about $V$ = 0) as well as Andreev reflection driven features (dips in $(dV/dI)_N$ symmetric about $V$ =0) in the spectrum. In this case the gap-structure appears at a lower voltage due to the thermal contribution to the spectrum\cite{NaidyukBook, GoutamPRB}. The magnetic field dependence of the same spectrum is also shown in Figure 2(C) where it is seen that the critical current ($I_c$) decreases with increasing magnetic field ($H$). The $I_c$ vs. $H$ plot for the same point-contact is shown in Figure 2(D). The magnetic field evolution of the gap structure (Figure 2(E)) shows that the gap changes almost linearly with magnetic field. The linear extrapolation of this dependence indicates a critical field of 4.5 T where the gap structure is expected to disappear.

Representative spectra obtained from three point-contacts in the ballistic regime are shown in Figure 3(A). The ballistic nature of such point-contacts were verified by measuring the temperature dependence of the normal state resistance that remained temperature independent. Two peaks symmetric about $V$ = 0 appear in the differential conductance ($(dI/dV)_N$) vs. $V$ spectra at 1.4 K -- this feature is a hallmark of Andreev reflection with a finite potential barrier at a superconducting junction\cite{BTK}. The peaks are significantly broader than what is expected from the theory of Blonder, Tinkham and Klapwijk (BTK) that is traditionally used to analyze Andreev reflection spectra obtained on conventional BCS superconductors\cite{BTK}. This might be due to a large inelastic broadening parameter at the interface\cite{broadening} or due to the existence of multiple gaps\cite{YNBC}. Nevertheless, the position of the peaks provide an approximate estimate of the gap \cite{BTK}. From the spectra provided in Figure 3(A) it is found that the magnitude of the gap in this case is approximately 6.5 meV. This value is unusually large given the low $T_c$ ($<$  6 K) of the superconducting phase as that points to a dramatically large value for $\Delta/k_BT_c \sim$ 10.

\begin{figure*}[!h]
\includegraphics[width=1\textwidth]{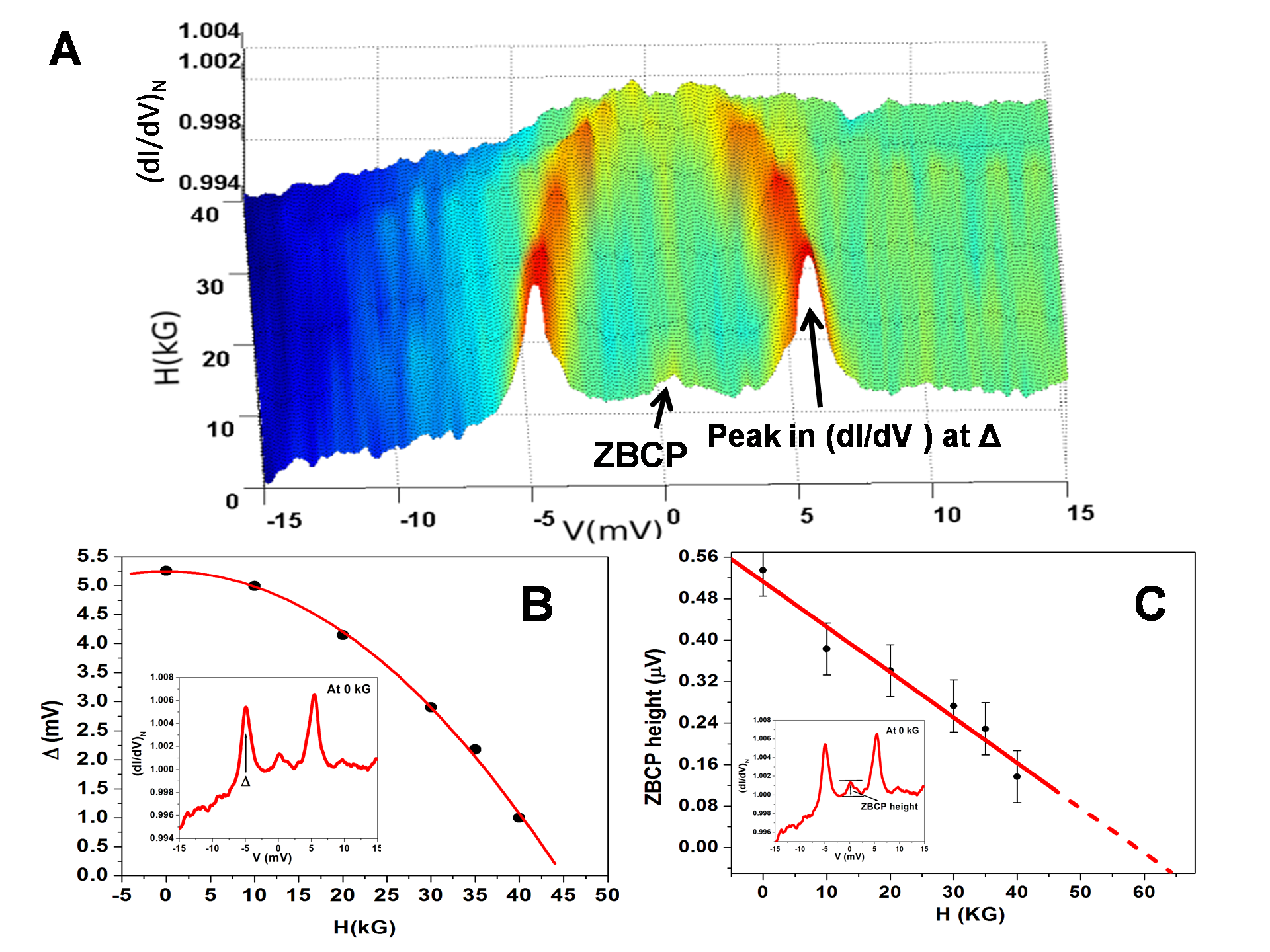}
\caption{(A) The magnetic field evolution of a spectrum in the ballistic regime showing the gap feature as well as a prominent ZBCP. (B) The field evolution of estimated $\Delta$. The inset shows the peak in $(dI/dV)_N$ corresponding to $\Delta$. (C) The field dependence of the ZBCP height. The scheme followed for mesuring the ZBCP height is shown in the inset.
 }
\end{figure*}
In order to obtain further insight about the gap feature observed at 1.4 K we have performed temperature dependence of the $(dI/dV)_N$ spectra in the ballistic regime. The temperature dependent spectra are shown in Figure 3(B). Surprisingly, the position of the peak in $(dI/dV)_N$  does not show significant shift with increasing temperature and the peaks survive up to 14 K beyond which the spectra become flat due to thermal broadening. This observation is strikingly similar to the pseudo-gap feature observed in case of the cuprates where the peaks in $(dI/dV)_N$ representing the pseudogap do not shift with temperature\cite{cupratepseudo}. It should also be noted that although the onset temperature of superconductivity is 6 K, the Andreev reflection like features are observed up to 13 K. Similar observation was made earlier in the context of the ferro-pnictide\cite{GoutamPRL, Laura2} and the chalcogenide superconductors\cite{Laura2}. This was attributed to (1) Andreev reflection originating from pre-formed phase incoherent Cooper pairs\cite{Choi} in the normal state of epitaxial thin films of ferropnictide superconductors\cite{GoutamPRL} and (2)a novel phase of matter not related to superconductivity in the normal state of the iron chalcogenide superconductors\cite{Laura2}. However, in the second case the spectral features exhibit a systematic temperature dependence unlike what we observe here. 

We should also consider two important facts here: (1) The spectra that is obtained in the zero-resistance state of the superconductor (at T = 1.4 K) does not alter at higher temperatures and (2) the temperature dependence shows striking similarities with what was observed for the pseudogap on the cuprates. Based on these it is rational to conclude that in the present case the peaks in $(dI/dV)_N$ originate from a normal state pseudogap of the superconducting phase formed at the constriction. 


In order to investigate the relation of this gap with superconductivity, we have probed the magnetic field dependence of the spectra obtained on another point-contact in the ballistic regime (Figure 4(A)). The gap features show systematic dependence on magnetic field and the extrapolation of this dependence indicates that the gap feature vanishes at 6.5 T. Therefore, it is confirmed that in this case the pseudogap is a precursor phase to the superconducting state.


All the spectra obtained in the ballistic regime also show a small peak structure in $(dI/dV)_N$ at $V$ =0. When the over-all signal is large, this structure is clearly visible. As shown in Figure 4(A), for certain point-contacts, this zero-bias conductance peak (ZBCP) is pronounced and clearly visible. The magnetic field dependence of this feature is presented in Figure 4(C). It is observed that the ZBCP does not undergo splitting and slowly fades away with increasing magnetic field. The linear extrapolation of the magnetic field dependence indicates that the ZBCP vanished at 6.5 T, which is same as the field where the other superconductivity related features disappear. Such a ZBCP with similar magnetic field dependence was earlier observed in the topological superconductor Cu$_x$Bi$_2$Se$_3$, where the ZBCP was attributed to the existence of Majorana Fermions\cite{SasakiPRL}. Therefore, our study points to the possible existence of Majorana Fermions in the superconductor formed at the point-contacts indicating the possibility of the new superconducting phase being a topological superconductor. 


In conclusion, we have discovered an exotic unconventional superconducting phase at nano-meter scale interfaces between pure elemental metals and the 3-D Dirac semi-metal Cd$_3$As$_2$ up to 7K. We also found evidence of a robust, temperature independent pseudogap up to 13 K. We found the signature of a ZBCP that could be the signature of topological superconductivity. The new phenomenon reported here is intriguing and therefore, this will motivate rigorous theoretical work. It is expected that following this work new devices involving the new 3-D topological Dirac semi-metals will be fabricated for exploring further intriguing phases of matter in confined dimensions. The superconducting phase presented here may be realized in nanostructured devices, interfaces, nano-particles and nano-composites involving Cd$_3$As$_2$ and elemental normal metals like Ag, Au or Pt.

We acknowledge professors S. Ramakrishnan, P. Raychaudhuri and I. Mazin for fruitful discussion. We would like to thank Dr. Deepanjan Chakraborty, Yogesh Singh and Sanjeev Kumar for critically reading the manuscript. We acknowledge the contribution of Mr. Avtar Singh, Ms. Preetha Saha and Ms. Shubhra Jyotsna. GS would like to acknowledge partial financial support from the research grant of Ramanujan fellowship awarded by the department of science and technology (DST), Govt. of India.

\


\begin{thebibliography}{99}

\bibitem{LiuNa3Bi} Z. K. Liu $et. al.$, Science \textbf{343}, 864 (2014).



\bibitem{LiuCd3As2} Z. K. Liu $et. al.$, Nature Materials \textbf{13}, 677 (2014).

\bibitem{YoungPRL} S. M. Young, S. Zaheer, J. C. Y. Teo, C. L. Kane, E. J. Mele, and A. M. Rappe, Phys. Rev. Lett. \textbf{108}, 140405 (2012).

\bibitem{WangPRB} Z. Wang, Y. Sun, X.-Q. Chen, C. Franchini, G. Xu, H. Weng, X. Dai and Z. Fang,  Phys. Rev. B \textbf{85}, 195320 (2012).


\bibitem{TI} M. Z. Hasan, C. L. Kane, Rev. Mod. Phys. \textbf{82}, 3045 (2010).


\bibitem{BurkovPRL} A. A. Burkov and Leon Balents, Phys. Rev. Lett.\textbf{107}, 127205 (2011)	. 

\bibitem{Waniridate} X. Wan, A. M. Turner, A. Vishwanath, and S. Y. Savrasov, Phys. Rev. B \textbf{83}, 205101 (2011). 

\bibitem{TSReview} X. -L. Qi, S. -C. Zhang, Rev. Mod. Phys. \textbf{83}, 1057 (2011).

\bibitem{TSPRL} L. Fu, C. Kane, Phys. Rev. Lett. \textbf{100}, 096407 (2008).

\bibitem{WangPRB2} Z. Wang, H. Weng, Q. Wu, X. Dai and Z. Fang, Phys. Rev. B \textbf{88}, 125427 (2013). 

\bibitem{QiPRB} X.-L. Qi, T.L. Hughes, and S.-C. Zhang,  Phys. Rev. B \textbf{81}, 134508 (2010). 


\bibitem{YazdaniSTM} S. Jeon et al., Nature Materials \textbf{13}, 851 (2014).
\bibitem{Supplementary} The detailed description of the experimental technique has been provided in supplementary materials. The synthesis and characterization of the samples and the analysis details are also provided. 

\bibitem{Wexler} A. Wexler, Proc. Phys. Soc. \textbf{89}, 927 (1966).

\bibitem{SupriyoDatta} S. Datta, Electronic Transport in Mesoscopic Systems, Cambridge University Press.

\bibitem{GoutamPRB} G. Sheet, S. Mukhopadhyay and P. Raychaudhuri, Phys. Rev. B \textbf{69}, 134507 (2004).

\bibitem{NaidyukBook} Yu. G. Naidyuk, I. K. Yanson, Point-contact Spectroscopy, Springer Newyork.
\bibitem{contact} The zero-resistance has not been measured here as the pseudo-four probe method employed here is not completely free from the contact resistances. The resistance measured at low temperature where the transition has been completed may be taken as the total contact resistance and when that is subtracted from the total resistance, the zero resistance can be obtained.

\bibitem{critical field} As expected, the critical field is also dependent on the size, shape and degree of disorder in superconducting point-contacts and therefore, $H_c$ is different for different point-contacts. The measured value of $H_c$ varied between 2 T and 3.5 T. 

\bibitem{BTK} G. E. Blonder, M. Tinkham and T. M. Klapwijk, Phys. Rev. B \textbf{25}, 4515 (1982).

\bibitem{broadening} A. Placenick {\textit{et al.}}, Phys. Rev. B \textbf{49}, 10016 (1994).

\bibitem{YNBC} P. Rayshudhuri, D. Jaiswal-Nagar, Goutam Sheet, S. Ramakrishnan, H. Takeya, Phys. Rev. Lett. \textbf{93}, 156802 (2004).

\bibitem{cupratepseudo} Ch. Renner, B. Revaz, J.-Y. Genoud, K. Kadowaki, and O. Fischer, Phys. rev. Lett. \textbf{80}, 149 (1998). 

\bibitem{Choi} H. Y. Choi, Y. bang, and D. K. Campbell, Phys. Rev. B \textbf{61}, 9748 (2000).

\bibitem{GoutamPRL} Goutam Sheet et.al, Phys. Rev. Lett. \textbf{105}, 167003 (2010).


\bibitem{Laura2} H. Z. Arham et al., Journal of Physics: Conference Series \textbf{400}, 022001 (2012).

\bibitem{SasakiPRL} S. Sasaki, M. Kriener, K. Segawa, K. yada, Y. Tanaka, M. Sato and Y. Ando, Phys. Rev. Lett. \textbf{107}, 217001 (2011).\\\\


 
\pagebreak
\renewcommand{\thefigure}{S\arabic{figure}}
\renewcommand{\thetable}{S\arabic{table}}
\setcounter{figure}{0}
\setcounter{table}{0}
\begin{center}
{\fontsize{50}{50}\textbf{\underline{Supplementary Materials}}}
\end{center}

\vspace{1 cm}





\begin{center}
\textbf{\underline{Synthesis}} 
\end{center}

Polycrystalline samples of Cd$_3$As$_2$ were obtained by heating the stoichiometric mixture of the constituent elements. A mixture of Cd and As powder ($\sim$ 1 gram) was sealed in an evacuated quartz tube ($\sim$ 10$^{-5}$ mbar), heated at 500$^{0}$C for 8 hours, then at 850$^{0}$C for 24 hours with a typical ramping rate of 1$^{0}$C/min and furnace cooled to room temperature. The shiny black crystalline product thus obtained was ground well, pelletized ($\phi$ = 8 mm) and heated again in vacuum at 400$^{0}$C for 6 h for homogenization. The pellet was shiny black and hard in nature.\\\\
\begin{center}
\textbf{\underline{Characterization}} 
\end{center}
\textbf{X-ray diffraction:} The samples were characterized by powder X-ray diffraction technique using Cu-K$\alpha$ radiation ($\lambda = 1.5406 Å$) on a Bruker D8 Advance diffractometer. All the peaks could be indexed on the basis of a centrosymmetric tetragonal cell in $I$4$_{1}$/acd space group as reported by Cava \textit{et. al.}\cite{Cava}. The sample was pure with no apparent impurity phase present in the resolution limit of X-ray diffraction analysis. Lattice parameters calculated using $Le`Bail$ method were in close agreement with the literature values (Fig. S1). In fig.S1 we show the $Le`Bail$ fit to the powder x-ray diffraction pattern of polycrystalline Cd$_3$As$_2$. The vertical bars indicate the allowed Bragg reflections.\\

\begin{figure*}
 \includegraphics[width=1\textwidth]{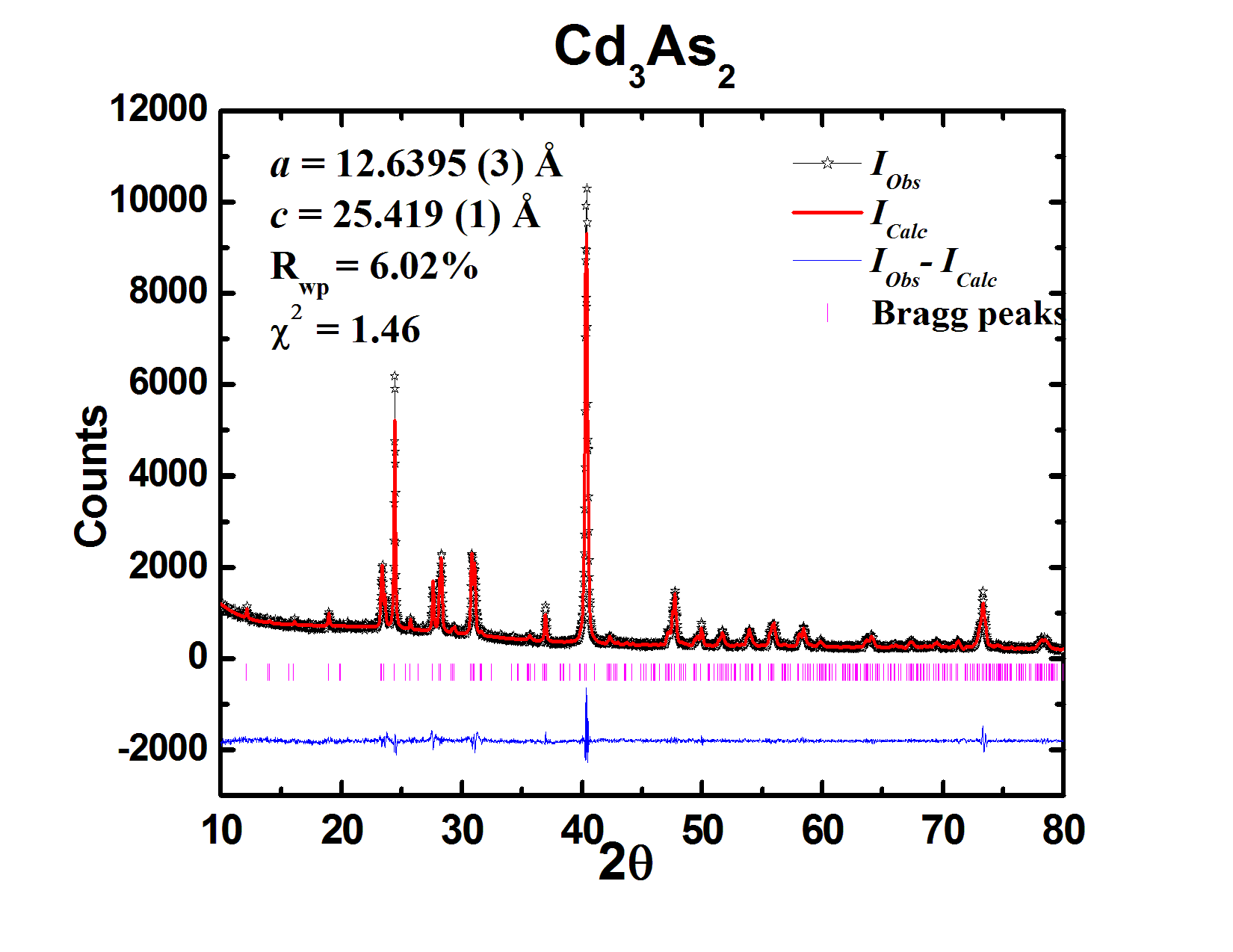}
 \caption{Powder x-ray diffraction pattern of polycrystalline Cd$_3$As$_2$ fitted with $Le`Bail$. The vertical bars indicate the allowed Bragg reflections.}
 \end{figure*}
 
 
 
 
\textbf{Energy dispersive X-ray analysis (EDAX):} Compositional analysis was done using a SEM-EDAX. The average stoichiometry found after collecting data on each sample at many different regions was close to 3:2 (Cd:As). There were some regions were As was slightly deficient ($~5\%$). Fig. S2 shows the presence of Cd and As. C and Si comes from the carbon tape and detector respectively. Inset of Fig. S2 shows a typical electron image of the polished pellet on which measurements were performed.\\\\

\begin{figure*}
\centering
\includegraphics[width=0.8\textwidth]{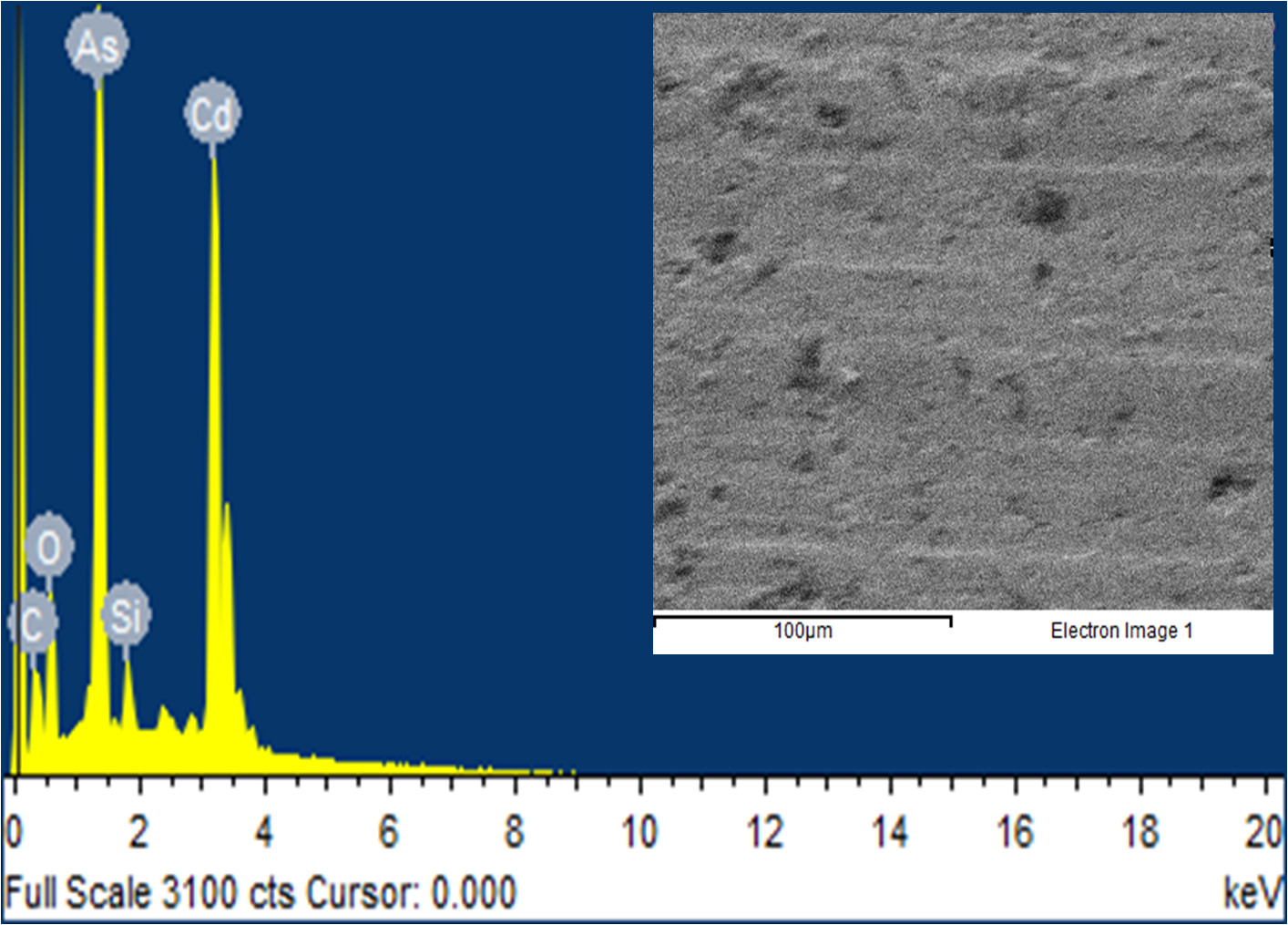}
\caption{ EDAX spectrum of Cd$_3$As$_2$. The inset shows image of polished pellet of Cd$_3$As$_2$.
 } 
\end{figure*}
 \vspace{.5 cm}
\textbf{Magnetization of the bulk Cd$_3$As$_2$ samples:} The VSM (Vibrating Sample Magnetometer) experiment was done to investigate the possibility of hidden bulk superconducting phase in the material. The measurement was done in a Quantum Design (QD) PPMS (Physical properties measurement system) with a VSM probe supplied by QD. The measurement field was 100 Oe. The magnetization does not show any diamagnetic transition down to 2 K. 

\begin{figure*}
\includegraphics[width=0.5\textwidth]{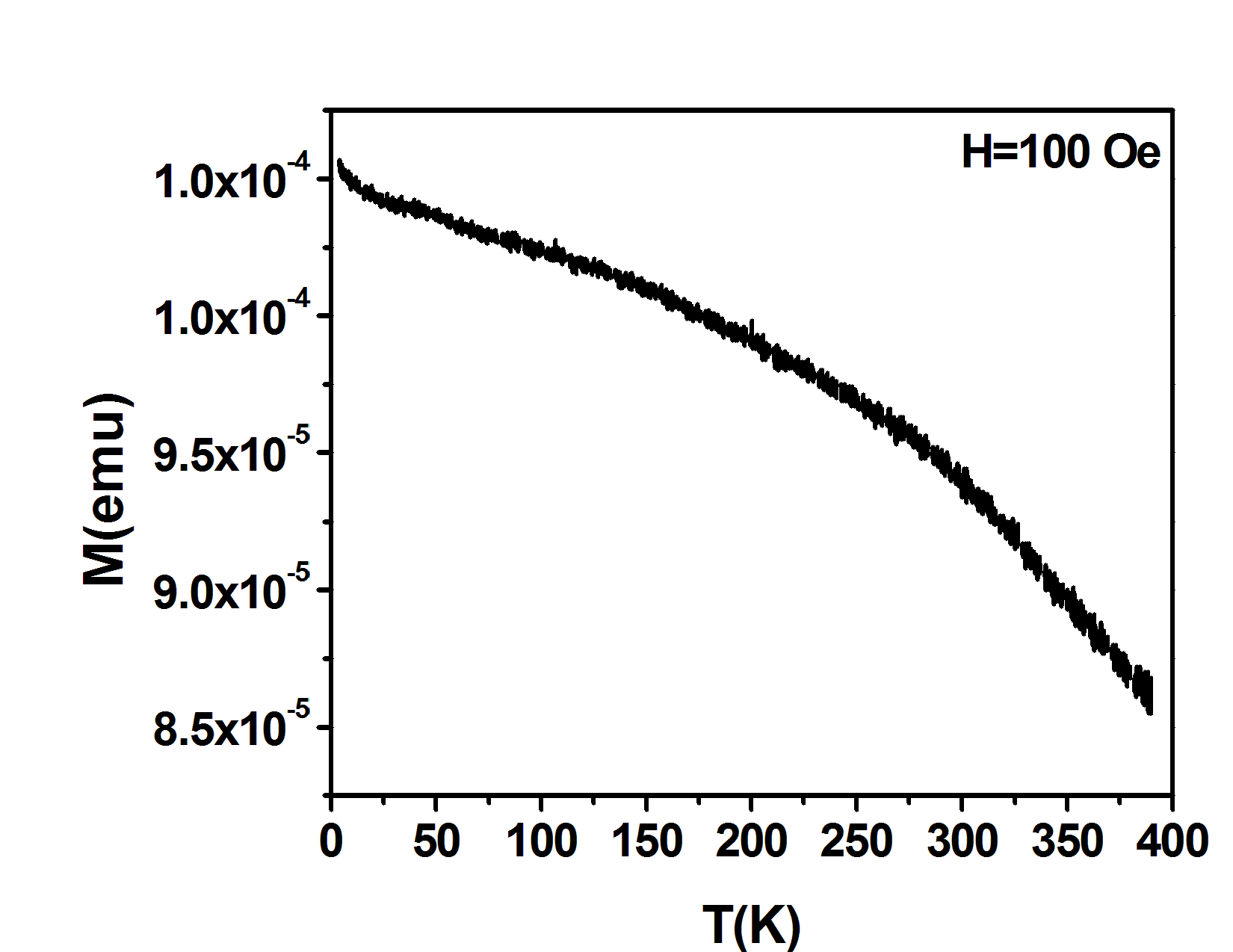}
\caption{ Magnetization vs. temperature of the bulk sample measured in a VSM in a Quantum Design PPMS. The bulk sample does not show any diamagnetic transition at low temperature down to 2 K. }
\end{figure*}

\textbf{Four-probe resistivity of bulk Cd$_3$As$_2$ samples:} Resistivity of the samples that were used for the point-contact spectroscopy measurements was measured by a four-probe technique. The resistivity as a function of temperature of one of the samples is presented in Fig. S4. The sample shows semi-metallic behavior down to 1.4 K. This further confirms that no superconducting phase is present in the bulk. 
\begin{figure*}
\includegraphics[width=0.5\textwidth]{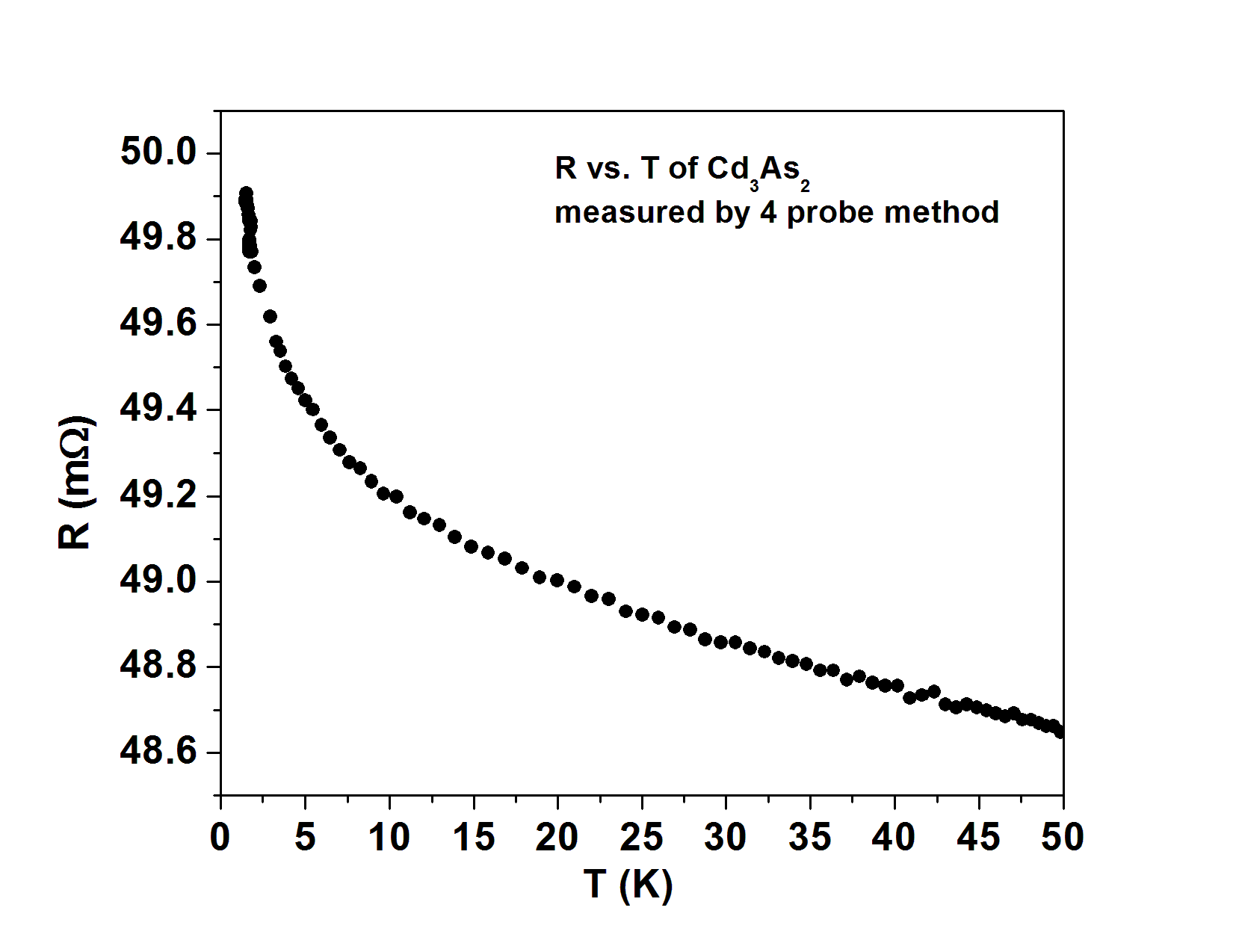}
\caption{ Four probe resistance vs. temperature of bulk Cd$_3$As$_2$.  }
\end{figure*}



 \vspace{1 cm}
 



\begin{center}
\textbf{\underline{Low-temperature measurements}} 
\end{center}

The low temperature measurements were performed in a liquid helium cryostat working down to 1.4K. The cryostat is equipped with a dynamic variable temperature insert (VTI) inside which there is one static VTI. The bottom part of the static VTI is made of copper for efficient cooling. The sample goes inside the static VTI which is first evacuated and then filled with dry helium exchange gas. The cryostat is also equipped with a three-axis vector magnet. The vector magnet can apply a maximum magnetic field of 6T along the vertical direction using a superconducting solenoid and 1T in the horizontal plane using four superconducting Helmholtz coils. For the measurements presented in this paper, magnetic field was applied in the vertical direction, perpendicular to the sample surface using the solenoid.

\textbf{Point-contact Spectroscopy:} Point-contact spectroscopy experiments were performed using a home-built low-temperature probe. The probe consists of a long stainless steel tube at the end of which the probe-head is mounted. The probe head is equipped with a 100 threads per inch (t.p.i.) differential screw that is rotated by a shaft running to the top of the cryostat. The screw drives a tip-holder up and down with respect to the sample. The sample-holder is made of a 1" dia. copper disk. A cernox thermometer was mounted on the copper disc for the measurement of the temperature. The temperature of the disc was varied by a heater mounted on the same copper disc.
The tips were fabricated by cutting a 0.25 mm dia. metal wire at an angle. The tip was mounted on the tip holder and two gold contact leads were made on the tip with silver epoxy. The samples were mounted on the sample holder and two silver-epoxy contact leads were mounted on the sample as well. These four leads were used to measure the differential resistance ($dV/dI$) across the point-contacts. The leads $3$ and $4$ were used to carry out the two-probe resitivity measurements.


\begin{figure*}
\includegraphics[width=0.8\textwidth]{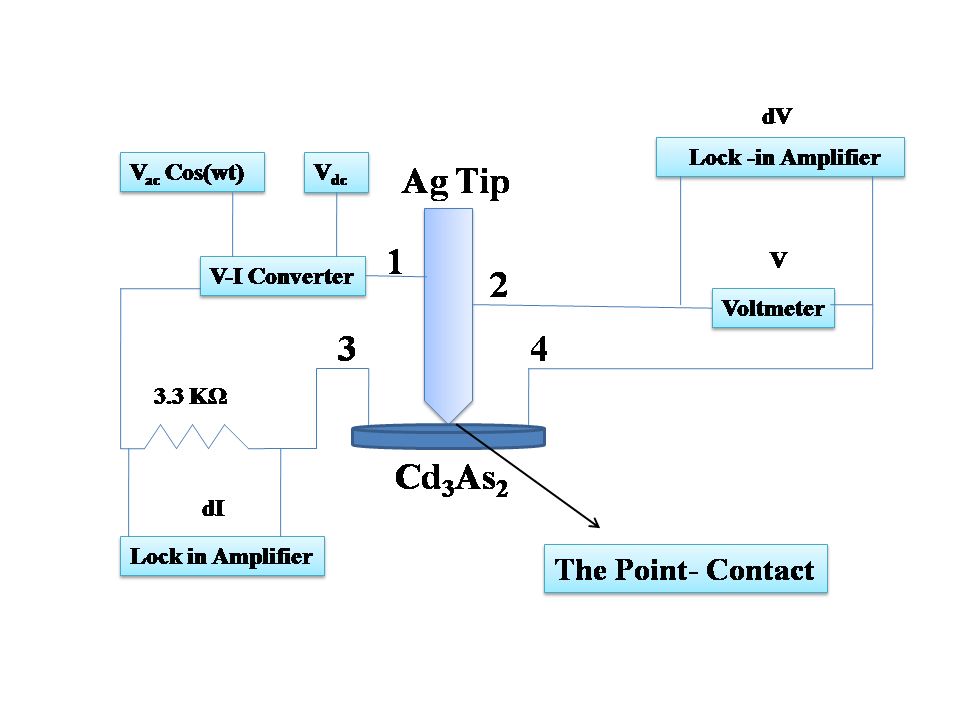}
\caption{ Schematic diagram describing the point-contact spectroscopy measurements.}
\end{figure*}

\vspace{.5 cm}
 
The point-contact spectra were captured by ac-modulation technique using a lock-in-amplifier (Model: SR830 DSP) (as shown in the schematic diagram). A voltage to current converter was fabricated to which a dc input coupled with a very small ac input was fed. The output current had a dc and a small ac component. This current passed through the point-contact. The dc output voltage across the point-contact, $V$ was measured by a digital multimeter (model: Keithley 2000 ) and the ac output voltage was measured by a lock-in amplifier working at 670 Hz. The first harmonic response of the lock–in could be taken to be proportional to the differential change in the voltage $dV$. The current was passed through a standard 3.3 $k\Omega$ resistor in series with the point contact. The drop across this resistor was measured by another lock in amplifier. This gave the estimate of the ac current passing through the point contact which is proportional to $dI$. $dI/dV$ is plotted against $V$ to generate the point-contact spectrum. We have normalized the spectra that are presented in this paper. The software for data acquisition was developed in house using lab-view.

\vspace{1 cm}

\textbf{How did we determine the critical temperature ($T_c$)?}



For all the point-contacts reported here we have measured the temperature ($T$) dependence of the point-contact resistance ($R$) with $V$ = 0. The $R-T$ data show a broad transition to the superconducting state. We have drawn the slope of the $R-T$ curves above and below the onset of transitions. The temperature at which the two slopes for a given $R-T$ curve meet has been taken as the $T_c$ for the corresponding point-contacts. It is important to note that for the ballistic point-contacts we cannot measure the $T_c$ as for such point-contacts the contact-resistance depends only on fundamental constants and remain temperature independent. However, from the thermal limit point-contacts we learn that the $T_c$ does not have a strong dependence on contact size and therefore it is rational to conclude that for the ballistic point-contacts $T_c$ remains close to 6 K.  



\paragraph*{} 
\textbf{Measurements with Pt and Au tips:} In order to investigate whether the new superconducting phase emerges only in a point-contact with silver (Ag) tip, we have performed measurements with platinum (Pt) and gold (Au) tips as well. It is observed that for all metallic tips the superconducting phase emerges. The data obtained between Cd$_3$As$_2$ and Pt and Au tips are shown in Fig. S6.

\begin{figure*}
\includegraphics[width=1\textwidth]{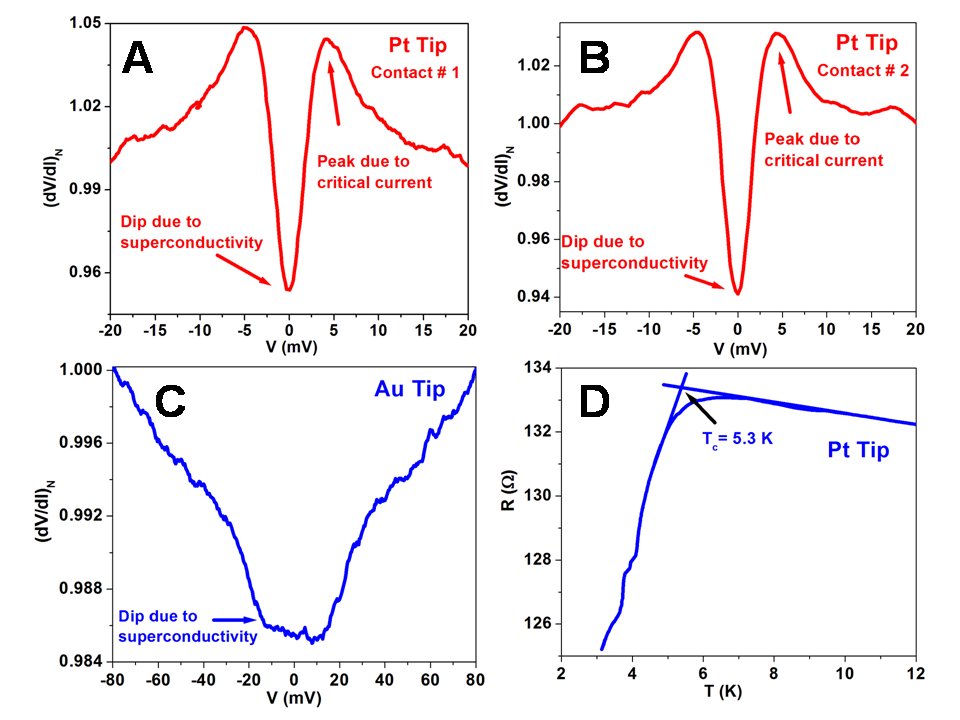}

\caption{ (A) and (B) showing representative spectra with superconducting dips that were obtained from two different point-contacts on Cd$_3$As$_2$ with a Pt tip.(C) A spectrum obtained with Au tip. This spectrum also shows the signature of the gap structure. (D) Resistance vs. temperature of the point-contact shown in (A). The superconducting transition is clearly seen at 5.3 K.  }
\end{figure*}



\textbf{On the size of the point-contacts:} The point-contact spectroscopy measurements were done at different points on two samples grown in two different batches. The contact size was estimated from the normal state resistance (at high $V$) following Wexler's formula given by:
\begin{displaymath}
                       R_{PC} = \frac{2h/e^2}{(ak_F)^2} + \Gamma (l/a)\frac{\rho (T)}{2a}
\end{displaymath}
          
where, $\Gamma(l/a)$ is a numerical factor close to unity. $a$ is the contact diameter and $2h/e^2$ is quantum resistance i.e. 50$k\Omega$. $k_{F}$ is the magnitude of the Fermi wave vector which is 0.04\AA~ for Cd$_3$As$_2$ and $\rho (T)$ is resistivity at temperature T(in K). 
$\rho (1.5) =  28 \mu \Omega-\rm{cm}$ for Cd$_3$As$_2$ (measured by conventional four-probe method).\\

\begin{figure*}
\includegraphics[width=.6\textwidth]{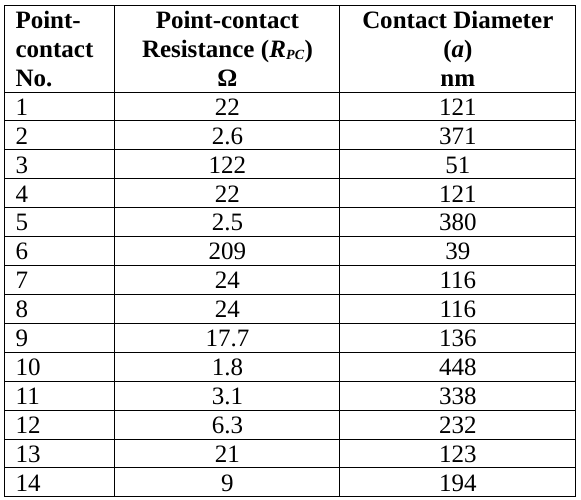}
\caption{ A representative list of the normal state resistance and size of the point-contacts. }
\end{figure*}

\vspace{1.5 cm}
\textbf{References:}
\bibitem{Cava}M. N. Ali, Q. Gibson, S. Jeon, B. B. Zhou, A Yazdani, and R. J. Cava,  Inorg. Chem.\textbf{ 53}, 4062 (2014).







\end{thebibliography}
\end{document}